# Impact of Review Valence and Perceived Uncertainty on Purchase of Time-Constrained and Discounted Search Goods

Prathamesh Muzumdar
pmmuzumdar@usf.edu
School of Information Systems & Management
Muma College of Business
University of South Florida
Tampa, FL

## Abstract

In the past decade, e-commerce industry has become a common source of electronic word of mouth (eWOM) for various products. Increasing online shoppers have generated enormous amount of data in form of reviews (text) and sales data. Aggregate reviews in form of rating (stars) have become noticeable indicators of product quality and vendor performance to prospective consumers at first sight. Consumers subjected to product discount deadlines search for ways in which they could evaluate product and vendor service using a comprehensible benchmark. Considering the effect of time pressure on consumers, aggregate reviews, known as review valence, become a viable indicator of product quality. This study investigates how purchase decisions for new products are affected by past customer aggregate ratings when a soon-to-expire discount is being offered. We examine the role that a consumer's attitude towards review valence (RV) plays as an antecedent to that consumer's reliance on RV in a purchase decision for time-discounted search goods. Considering review credibility, diagnosticity, and effectiveness as determinants of consumer attitude in a time-constrained search and purchase environment, we follow the approach-avoidance conflict theory to examine the role of review valence and perceived uncertainty in a time-constrained environment. The data was collected through an online survey and analyzed using structural equation modelling. This study provides significant implications for practitioners as they can better understand how review valence can influence a purchase decision. Empirical analysis includes two contributions: 1. It helps to understand how consumer attitude toward review valence, when positively influenced by the determinants, can lead to reliance on review valence, further influencing purchase decision; 2. Time constrained purchase-related perceived uncertainty negatively moderates the relationship between consumer attitude and reliance on review valence.

Keywords: Online Consumer Reviews, Review Valence, Perceived Uncertainty.

## 1. INTRODUCTION

In the last decade, online customer reviews (OCRs) have emerged as an important source of information for prospective buyers, substituting other forms of marketing promotions. OCRs act as electronic word-of-mouth (eWOM) for buyers (Q. B. Liu & Karahanna, 2017). OCRs are significant indicators of product quality, reliability, and performance (C. Liu & Forsythe, 2010). The advantage of OCRs is their accessibility compared to other forms of WOM and marketing promotions. Consumers can make their opinions easily accessible to other consumers through the Internet (Z. Zhang et al., 2020). The literature on eWOM has shown that the OCRs significantly influence customer purchase behavior (Q. B. Liu & Karahanna, 2017; C. Liu & Forsythe, 2010). further influencing product sales (Q. B. Liu &

Considering the influence of information available through various digital forms, it is important to understand the effects of time pressure and product promotions on consumers' incorporation of such information. In this study, we account for the significance of time pressure relating to a consumer purchase decision. Goods can be classified across a continuum of search, experience, and credence claims. We will not consider credence goods in this paper (Z. Zhang et al., 2020). Experience goods can only be accurately evaluated after the product is purchased and then used (Z. Zhang et al., 2020). Search goods are both non-experience and non-credence goods that are evaluated prior to purchase using prior knowledge, direct product inspection, reasonable effort, and normal channels of information acquisition, such as Consumer Reports (Ford et al., 2021). Such goods have discounted prices during promotions, thereby being time-constrained for purchase. This generates a complex conflict for prospective consumers in making purchase decisions. Relying on few evaluation parameters, prospective consumers seek a shorter alternative to longer OCRs to make a decision.

Time-discounted search goods are non-experience goods which have discounted price for a specific time frame. This makes it deceptive for consumers to make an uncertain purchase decision for these products. Time constraints leave them with few options to evaluate the quality and performance unfamiliar product. Review valence plays a pivotal role in helping consumers unearth the insights in its quantified shorter evaluation form, e.g., product star rating plus number of reviews (Wang et al., 2020). Thus, consumers prefer to use aggregate reviews (review valence) as a measure to quickly judge product quality and performance to make the purchase decision (Allard et al., 2020). RV becomes a parameter to quickly judge a non-experience product and support the purchase decision; nevertheless, it still produces unforeseeable uncertainty among consumers. Most of the existing literature has focused on examining the influence of online consumer reviews on purchase decisions for experience goods (H. Zhang & Gong, 2020). In contrast, only a handful of studies have focused on search goods. Not much has been done to understand the role of OCRs on time-constrained price discounted search goods. To date, no study has examined the effects of review valence on purchase decisions for time-discounted search goods.

This study uses the approach-avoidance conflict theory to understand how time pressure influences consumers' evaluation parameters, affecting the purchase decision. Approach-avoidance conflict theory suggests that conflicts occur when a specific event or goal has appealing and unappealing characteristics (Penz & Hogg, 2011). Discounted search goods having purchase time pressure might lead to conflicts in the form of low price (appealing) or bad quality (unappealing). These outcomes are related to time pressure, which compels the consumer to make a quick decision based on a few easily comprehendible parameters like review valence (aggregate ratings). We, as researchers, try to address the following questions in this study:

1. Under time pressure, what impact does review credibility, diagnosticity, and effectiveness have on consumers' attitude towards review valence (RV) while using discounts on search goods?
2. Under time pressure, how does consumers' attitude toward review valence influence their reliance on review valence for making a purchase decision on search goods?
3. Under time pressure is perceived uncertainty, a significant moderator of the consumer's attitude in making a review valence less relevant?

The remainder of this paper is organized as follows: Section 2 discusses the theoretical background of the study, which includes five sub-sections. Section 3 takes into account the conceptual model and hypotheses. We have collected the data using online surveys and then analyzed them using structural equation modelling (SEM). Section 4 presents the methods and measurements used in this study. Section 5 outlines study results, and section 6 discusses research findings. The paper concludes with a summary of research findings and implications for future research and practice in Section 7.

## 2. THEORETICAL CONTEXT

WOM, in general, is defined as an informal advice or communication about products, services, and brands that can be communicated from one customer to another in person or through a distance communication medium (Mandal et al., 2021). eWOM is electronic word of mouth that is digitally communicated through the Internet (Beurer-Zuellig & Klaas, 2020). Online consumer reviews are the most exclusive eWOM omnipresent in different forms on online retail outlets. Because online consumer reviews are initiated by customers independent of the market, they are perceived to be more reliable and trustworthy than other communications (Mandal et al., 2021). Mandal et.al (2021) showed that OCRs are widely accepted as eWOM and are closely related to business success.

The growth of online retail and the reach of the internet has allowed consumers to share their experiences using OCRs; this provides the consumers with an online channel to share their product evaluations. As a product of this process, online consumer reviews have emerged as a phenomenon influencing consumer purchase decisions (Tonietto & Barasch, 2020). Compared to traditional promotional marketing techniques and WOM, which are limited to a local physical social network (Lin & Xu, 2017), the impact of online consumer reviews is beyond local communities. It uses information technology and internet tools to reach people all over the world (Clemons et al., 2006). In their study, Wang et.al (2020) inferred that traditional WOM generally does not play the role of a direct decision variable for product sales. Recent research by Jensen et.al (Jensen et al., 2013) found a direct connection between online consumer reviews (eWOM) and product sales. For example, Kim et.al (Kim et al., 2011) studied the effect of online consumer reviews on hotel bookings.

**Experience goods vs. time-discounted search goods**
Consumers subjectively evaluate experience goods through sampling or purchase in order to evaluate their quality (Calderón Urbina et al., 2021). On the other hand, search goods are evaluated by feature properties, and consumers usually do not require interacting with the product for evaluation (H. Zhang & Gong, 2020). Examples of experience goods include music, books, and soda. Search goods include smartphones, cameras, and clothing (Mandal et al., 2021). With the rise in online retail, all search and experience goods features are searchable, and the traditional distinction between experience goods and search goods has been reduced (Calderón Urbina et al., 2021). However, the research by Zhang et.al (2020) found that the distinction is still valid due to the different ways in which product-related information is accessed and processed. Their study also shows that online consumer reviews help make purchase decisions for search goods (Calderón Urbina et al., 2021; H. Zhang & Gong, 2020).

**Online consumer reviews and time-discounted search goods**
OCRs about technology products are considered more relevant to customers than online marketing promotional information created by sellers (Mandal et al., 2021). Promotional information mostly includes the product's technical specifications. For experience goods, this information is important and helps consumers relate the technical features to their experience (Calderón Urbina et al., 2021; H. Zhang & Gong, 2020). In contrast, product details are important for consumers for search goods but are not enough to support their purchase decision (Tonietto & Barasch, 2020). Zhang et.al (2020) showed that consumers are also interested in knowing how other consumers feel about the product, technical specifications, and product conclusion. Online consumer reviews are provided by consumers who have used the product for a certain period and know the product's features.

Consumers who find it difficult to form an opinion on product purchases use online consumer reviews to help them comprehend the benefits of such products (Calderón Urbina et al., 2021). Non-experience buying relies heavily on consumers' ability to form an opinion from the information in the reviews. Consumers highly rely on online consumer reviews to support their purchase decision (H. Zhang & Gong, 2020) in case of search goods. In this study, we examine the role of OCRs in shaping consumers' attitudes toward review valence, especially when deciding on the purchase of search goods.

**Approach-avoidance conflict theory**
Intertemporal choices are defined as decisions that have consequences in multiple periods (Penz & Hogg, 2011). These choices require decision-makers to trade-off costs and benefits at different points in time (C. Liu & Forsythe, 2010). A decision about cashing a discount for goods is an intertemporal choice. Descriptive discounting models capture the phenomenon that most economic agents prefer current rewards to delayed rewards having similar magnitude (Penz & Hogg, 2011; C. Liu & Forsythe, 2010). Most current rewards in the form of smaller rewards are considered immediate discounts (D. Zhang et al., 2016). In this study, we examine the effects of immediate discounts on consumers' ability to make quick purchase decisions. Offers having a limited validity time for participation may increase discount redemption in a shorter period if consumers know the expiration date. Pressure to take action before the offer ends or time pressure and information about the offer are components of persuasion. Consumers seek to avoid losses associated with missed opportunities by making quick purchase decisions based on few selected product evaluation factors (Lee & Hong, 2021). This study proposes approach-avoidance conflict theory as a plausible theoretical mechanism to discuss the effects of time pressure and discounts on purchase decisions.

Approach-avoidance conflict theory shows the duality of event outcomes that occur when events are appealing and non-appealing simultaneously (Clemons et al., 2006). RV is a very comprehendible measure to evaluate a product in a shorter span of time and helps support consumers in making their purchase based on that one criterion. Though that makes the process less arduous for consumers to jump on the purchase decision, it also develops unsettling perceptions among consumers. These perceptions lead to uncertainty among consumers on their reliance on RVs. Do RVs give us complete insights on the product on display? Is this the question those consumers have in their minds? The use of RVs occurs due to the time pressure that promotions build during product sales, and perceived uncertainty comes out to be an unappealing outcome of such decisions. This study considers both time pressure and approach-avoidance conflict theory to examine how uncertainty influences consumers' reliance on easily comprehendible product criteria like RV.

**Credibility and diagnosticity of OCRs**
The credibility of OCRs in eWOM literature has been studied extensively for several years (Jha & Shah, 2021). Credibility in communications literature is defined as the extent to which a communication source is considered valid and perceivable to the reader (Jha & Shah, 2021; Cheung et al., 2012). Some have defined credibility as evaluation done by readers concerning the believability of a reviewer (Cheung et al., 2012). Diagnosticity is defined as the adequacy of a piece of conclusive information provided to the reader about the relevance of the information to the judgmental task (Weathers et al., 2015). A review is evaluated for its diagnosticity by the relevancy of the information it provides to the actual task, which the reader wants to complete (Cheung et al., 2012). Information usefulness for making a judgment over a decision is what makes the information very relevant. Relevancy of the information in reviews leads to diagnosticity (Jha & Shah, 2021). This study on OCRs heavily relies on consumers' evaluation of reviews to help them understand the product features and performance. Thus, these two variables play important roles in helping consumers evaluate reviews and develop an attitude toward reviews.

**Review effectiveness**
The effectiveness of online communication is well studied in IS literature. Online reviews are a type of user-generated content (UGC); their effectiveness plays an important role in influencing the readers' decision. Review effectiveness is defined as the degree to which a review can help consumers comprehend information and understand the judgmental task (Beurer-Zuellig & Klaas, 2020). Review effectiveness is multi-dimensional, and its three dimensions are popularity, helpfulness, and persuasiveness (Lin & Xu, 2017; Wu, 2017). Likes on reviews denote the helpfulness of the review (Hu et al., 2008); they indicate the richness of the information contained in the review. Hu et.al (2008) showed that highly liked reviews represent the predisposition of a review in helping consumers evaluate the information contained in the review. Review popularity represents the proneness of a review in attracting consumer attention (Zou et al., 2011) and is responsible for building awareness among consumers (Lin & Xu, 2017). Review persuasiveness is the final determinant of effectiveness; it convinces consumers to persuade and committing to making purchases (Kuan et al., 2015). In this study, we examine the role of effectiveness in influencing consumers' attitudes toward RV.

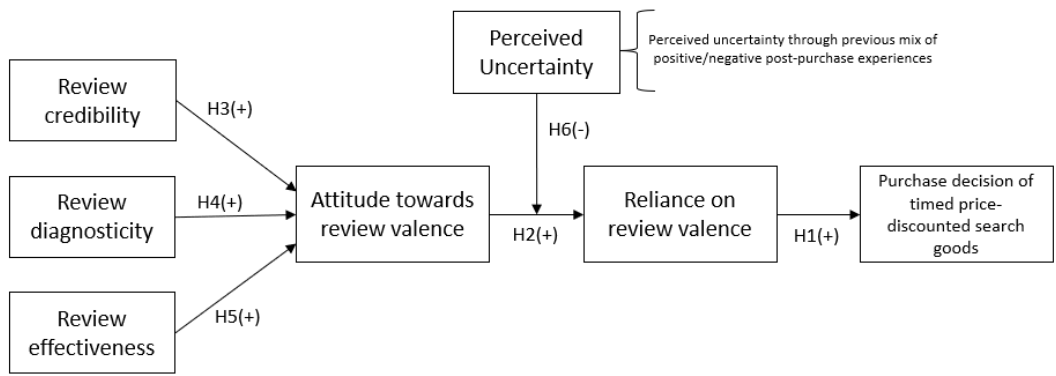

**Figure 1. Conceptual model for purchase decision of timed price-discounted search goods**

# 3. CONCEPTUAL MODEL AND HYPOTHESIS

OCRs driven by their attitudes towards reviews, which in turn are posited to be driven by review trustworthiness (credibility & diagnosticity) and effectiveness.

**Reliance on review valence and purchase decision**

In this study, reliance on review valence is viewed as the extent to which consumers depend on aggregate ratings to make their purchase decision. Reliance addresses the extent to which a consumer feels a need to use OCRs before making purchase decisions. At the same time, consumers worry about the decision quality if they do not adhere to extraneous advice through OCRs. Aggregate rating in the form of review valence becomes an easy way to assess product quality in a shorter time frame, helping consumers to decide for the purchase of timed price-discounted search goods. In the case of experience goods, previous experience with products makes it easy for the consumer to come up with the purchase. However, when it comes to timed price discounted search goods, it becomes important to purchase within the specified time frame to avail discount. For such consumers, it becomes important to rely on the aggregate ratings in the form of review valence. Consumer expertise can be expressed in the form of an online review (eWOM), helping new consumers get an insight into what the product has to offer consumers. However, more importantly, those insights in the form of ratings can help new consumers quickly conclude their decision. This conceptualization of online aggregated ratings (review valence) draws on the conclusion that reliance on online review valence is a more complex construct than simply following eWOM and traditional WOM; the amount of time spent with the medium and more belief in aggregated rating determines the severity of the influence on the consumer (Allard et al., 2020; Tonietto & Barasch, 2020; East et al., 2007). Therefore, we hypothesize that:

H1: Consumers' purchase decision of timed price-discounted search goods is positively influenced by their reliance on review valence.

**Attitude toward review valence**

In this study, attitude is defined as a tendency to evaluate an opinion with some degree of favor or disfavor, usually expressed in cognitive and behavioral responses (Tonietto & Barasch, 2020). Attitude toward online review valence speaks about consumers' feelings about online consumer

The proposed conceptual model for the study is shown in Figure 1. Purchase decision is posited to be driven by consumers' reliance on reviews (Allard et al., 2020). A general tendency to view OCRs in either a positive or negative light gets reflected in their attitude. Consumers comprehend review valence better than online consumer reviews. Since it is an aggregated rating value, it helps consumers conclude in lesser time (East et al., 2007). This quickly leads to developing a positive or negative attitude toward review valence. This attitude further influences the consumers' reliance on review valence. Therefore, we hypothesize that

H2: Consumers' reliance on review valence is positively influenced by consumers' attitudes toward review valence.

**Credibility and diagnosticity of online consumer reviews**

In this study on OCRs, credibility is the extent to which consumers trust OCRs to deliver truthful and accurate product information. Diagnosticity signifies review relevancy towards the task at hand. Past studies have shown that credibility judgments and diagnosticity influence consumers' attitudes in various contexts (Jensen et al., 2013). In the case of OCRs, Zhang et.al (2016) found that along with time spent on a retailer website and product specifications, OCR credibility is an important determinant of attitude toward the OCR. Kaun et.al (2015) showed that consumers rely heavily on diagnosticity to believe in the facts presented in the information, and somewhere this affects their attitude toward the review. As time plays a crucial role in time-discounted search goods purchase decision, it is very important to examine how consumers perceive review valence (aggregate rating) as an accurate indicator of product evaluation. Shorter time makes it taxing and tenuous for consumers to read and appraise all reviews. Review valence becomes a strong indicator of product performance and quality at first glance, followed by reading selective reviews to support their thoughts on review valence. This study is determined to explore the role of review credibility and diagnosticity in the context of time-discounted search goods, wherein the shorter time frame to cash in the discount makes it difficult for the consumer to spend more time reading reviews. Therefore, we hypothesize that:

H3: Consumers' attitude towards review valence is positively influenced by the perceived review credibility of OCRs.

Review diagnosticity is defined as the degree to which a consumer can rely on reviews to make purchase decision (Chua & Banerjee, 2014). In this research, review diagnosticity is associated with review depth and review readability.

H4: Consumers' attitude toward review valence is positively influenced by the perceived review diagnosticity of OCRs.

**Review effectiveness**
Previous research on effectiveness has heavily focused on understanding its determinants and its effects on purchase intention. In their study, Lin et.al (2017) showed that review effectiveness is a determining factor of consumers' persuasion of a product. It influences the consumers' attitude toward OCRs by trusting the information in the review (Wu, 2017). Review persuasiveness is considered one of the determinants of effectiveness, influencing the consumers' overall attitude toward review information (Hu et al., 2008). Review helpfulness exhibits the richness of the information and its relevancy toward product features (Wu, 2017). Review popularity attracts consumers toward reviews and makes them more prone to believing in the information in the review (Cheung et al., 2012). Time pressure and price promotions make it tedious for consumers to read every review posted in support or against the product. It is important to understand the role of effectiveness on time-constrained evaluation criteria like review valence. Therefore, we hypothesize that

H5: Consumers' attitude toward review valence is positively influenced by the perceived review effectiveness of OCRs.

**Perceived uncertainty**
Product uncertainty refers to a situation where consumers realize at the post-purchase stage that the product, they bought is different from what they perceived it to be at the shopping stage. Such experiences lead to uncertain decisions during search and purchase periods. Perceived uncertainty is defined as emotional costs associated with unexpected losses that could occur after purchasing the product, caused by information asymmetry (Lee & Hong, 2021). The goal of the consumer is to evaluate the intrinsic quality of a product based on the information available in the reviews and then purchase the product with the lowest uncertainty. Search goods are non-experience goods; most consumers may or may not have previously used the product or conducted business with the online vendors. In such cases, there are financial and psychological uncertainties associated with the product and online vendors (Hong et al., 2017).

According to approach-avoidance conflict theory, events can have appealing and non-appealing outcomes. In such cases, perceived uncertainty can generate the fear of unexpected losses due to non-appealing outcomes. To understand the effects of perceived uncertainty on consumers' purchase decisions, we examine its moderating effects on consumers' attitudes and reliance on RV. As time pressure plays a crucial part in cashing discounts, we hypothesize that

H6: Perceived uncertainty negatively moderates the relationship between attitude toward review valence and reliance on review valence.

### 4. METHODS

The data to test the hypothesis was collected through a self-administered structured online survey using respondents drawn from Survey Monkey's panel of US consumers. Responses were collected only from respondents who had read or used an OCR within the past six months for searching for goods which they never experienced or used before. It was made sure through screening section that respondents were looking for price discounted goods with time deadline. A sample of 320 responses was purchased, and the sample size was established based on the guidelines in the SEM literature. The sample sizes are recommended to be between 100 and 400 respondents for the simple SEM model used in this study. It helps avoid unstable solutions at low sample sizes and sensitivity issues at large sample sizes (>500), often resulting in poor model fitting.

**Measures and measure validation**

All items for reliance on review valence and attitude towards review valence were adapted from Zou et.al (Zou et al., 2011) except items 3 and 4 from the variable attitude toward review valence. Items for reliance on RV reflect different dimensions of reliance as captured in dictionary definitions. In contrast, items for attitude toward RV reflect the degree of positivity or negativity that a consumer has toward RV in general. All items for review credibility and diagnosticity were adapted from Ghazisaeedi et.al (2012) and Hennig-Thurau and Walsh (2003), with adaptations made to reflect consumers' perceptions of the credibility and diagnosticity of OCRs. The items for review effectiveness and review uncertainty of OCRs were adapted from Kim et.al (Kim et al., 2011), which reflect the consumers' perception of the effectiveness of the information presented in the review. Items for perceived uncertainty reflect the unpredictability consumers feel when comprehending information from the reviews. Consumer's purchase decision was measured using a single question, it is also the dependent variable in the conceptual model. All the items were measured using a Likert-type scale to which respondents expressed agreement/ disagreement on a seven-point scale (1 = strongly disagree; 7 = strongly agree).

Following Anderson and Gerbing (1988), before conducting structural analysis for hypothesized relationships, the construct measures were validated through confirmatory factor analysis (CFA) using LISREL for Windows. Table 1 summarizes standardized factor loadings, composite reliability, average variance extracted, and Cronbach's alpha. All the items were retained as standardized factor loadings were above the recommended level of 0.5 (Anderson & Gerbing, 1988). Construct reliability was measured via composite reliability and Cronbach's alpha to estimate the consistency of the construct. The values for both the constructs in Table 1 exceeded the minimum threshold value of 0.70, signifying the high reliability of the constructs. Convergent validity was verified through average variance extracted (AVE); it measured the overall variance in the indicators as truly representative of the latent construct. The AVE values ranging from 0.660 to 0.872 implied that convergent validity was achieved because all items in the measurement model were statistically significant.

The overall model fit statistics (Table 1) show an acceptable fit of the measurement model to the data [$x^2$(288 df) = 308 ($p < 0.001$); Comparative Fit Index (CFI) = 0.98; Root Mean Square Error of Approximation (RMSEA) = 0.058; Goodness-of-Fit Index (GFI) = 0.90; Adjusted Goodness-of-Fit Index (AGFI) = 0.84]. RMSEA is just slightly higher than the recommended minimum value of 0.05, GFI is 0.94 (above 0.9 is preferable), and AGFI is slightly below 0.9 at 0.89. Table 2 shows discriminant validity; it was checked by comparing the shared variance among variables with the square root of AVE by each construct. The shared variances among factors are lower than the square root of AVE. We conclude that the discriminant validity was achieved.

**Common Method Bias**
To address common method bias we analyzed the data through Harman's single factor analysis using principal axis factoring (Jordan & Troth). For results we extract 30.8% of variance which is less than 50%. We conclude that no common method bias exists in our measurement.

## 5. ANALYSIS AND RESULTS

**Descriptive Statistics**
Table 3 shows the means and standard deviations for all the constructs. Means for all the constructs are above the scale mid-point of 4. One sample t-tests were conducted to test if one can conclude that scores of the constructs are above the scale mid-point in the larger population based on the sample means. The results show that all the t-values are statistically significant at the 1 percent level. Thus, we conclude that, in general, the study population finds OCRs to be both valuable and credible; they have positive attitudes towards review valence and generally rely on these aggregate ratings in product purchase decisions.

**Hypotheses Tests**
The hypothesized relationships (H1 to H4) were tested using structural equation modeling (SEM) (table 4) by adding structural parameters to the measurement model in Table 1. For this test, the structural model was run on the entire sample. The coefficient for the reliance on review valence and purchase decision relationship is positive and statistically significant ($b = 0.42$; $p < 0.01$). In general, consumers' reliance on review valence positively influences their purchase decision for timed discounted search goods, supporting H1. Aggregate reviews can be a strong determinant of purchase decisions. From table 4, the coefficient for the attitude toward review valence and reliance on review valence relationship is positive and statistically significant ($b = 0.68$; $p < 0.01$). It implies that a positive attitude towards review valence can lead to more reliance on review valence, supporting H2. The coefficients for review credibility ($b = 0.76$; $p < 0.01$), review diagnosticity ($b = 0.48$; $p < 0.01$), and review effectiveness ($b = 0.52$; $p < 0.01$) are positive and statistically significant, supporting hypotheses H3, H4, and H5, respectively. Thus, both factors are significant drivers of consumers' attitudes toward review valence. In relative terms, however, perceived credibility has a greater impact than perceived diagnosticity.

**Moderator effects**
A moderator analysis was performed in SEM to test the two moderators in table 5. The moderating effect of review effectiveness on the relationship between attitude toward review valence and reliance on review valence and perceived uncertainty on the relationship between reliance on review valence and purchase decision. A moderating effect is identified when the chi-square significantly increases after the paths are constrained. Table 5 shows the results of the moderating test for the overall model and each path. The chi-square change of the overall model is significant ($p<0.001$), showing a possible moderating effect and supporting H5 and H6. Thus, both aspects are determined to be significant drivers of consumers' attitudes toward review valence. In relative terms, however, review credibility has a more significant impact than review diagnosticity.

## 6. DISCUSSIONS AND IMPLICATIONS

This study examined the role of review credibility and review diagnosticity (OCRs) on consumers' attitudes toward review valence and how such attitudes impact the extent to which consumers rely on review valence in purchase decisions of non-experience goods when the decision is time-constrained. Results show that review credibility and review diagnosticity are strong positive drivers of attitudes toward review valence, with review credibility having a relatively higher impact. In turn, attitudes strongly predict the tendency to rely on review valence. Additional analyses show a significant moderating effect of review effectiveness and perceived uncertainty. It is also noteworthy that respondents found review valence (aggregate ratings) credible, relevant, and effective, as evidenced by the high mean scores. Respondents likewise had positive attitudes toward review valence and generally relied on the aggregate ratings for product purchase decisions. The results have theoretical and managerial implications.

### Theoretical Implications
From a theoretical point of view, this research adds to the OCR literature in two important ways. First, it introduces two constructs that can add to our understanding of how consumers relate to review valence (aggregate rating) when it comes to time-discounted search goods and how they rely on the aggregate rating of the review valence to support their purchase decision. The construct of reliance on review valence adequately captures a growing phenomenon that has been observed in many recent consumer surveys about time-constrained discounted goods, i.e., consumers reporting an increasing tendency to rely on review valence for many purchase decisions of time-constrained discounted goods, while overlooking most reviews in the process. Attitude toward review valence is a relevant construct in the digital economy and retail. There is a growing realization by consumers that OCRs are subjective regarding the credibility and diagnosticity of the reviews. The study calls attention to these two new constructs and provides initial conceptualizations and empirical analysis.

Second, this study contributes to the limited literature on the possibility of perceived uncertainty as a moderator, regulating the relationship between consumers' attitudes and reliance on review valence. While perceived uncertainty has been explored in different capacities in other studies (Clemons et al., 2006), this study considers the uncertainty developed under time-constrained circumstances while purchasing a time-discounted search goods. The present study found a significant impact of perceived uncertainty as a moderating variable. Furthermore, it was successful in providing theoretical and empirical grounds for expecting the existence of approach-avoidance conflict theory in OCRs. Further research is needed, possibly in different contexts, to understand the role of perceived uncertainty in moderating the main effects of OCRs.

### Industrial Implications
From an industrial point of view, this study's findings are helpful for marketing managers to the extent that they demonstrate the power that review valence (aggregate reviews), a very common noticeable value in OCRs, exert on consumer purchase decisions. The findings also suggest that managers also need to recognize the importance of perceived uncertainty in moderating the relationship between attitude and reliance. Furthermore, the study signifies the importance of review credibility in driving the attitude and further reliance. Thus, the online review system needs to employ techniques that help reduce uncertainty and generate review credibility in time-constrained environments, e.g., by finding ways to communicate the expertise (knowledge) and trustworthiness (unbiased motives) of reviewers (Weathers et al., 2015).

## 7. LIMITATIONS AND SUGGESTIONS FOR FUTURE WORK

This study has limitations that future studies could address. First, it focused on the effects of review valence on time-discounted search goods. However, given the widely held notion that higher review volume could suffice the genuineness of OCRs, it is essential to study the effect of the volume for such goods. Does volume moderate the relationship between attitude and reliance? At lower volumes, do consumers rely on higher aggregate ratings? If yes, what are the determinants of such phenomena?

Second, the moderating effects of perceived uncertainty exist when we consider the approach-avoidance conflict theory. It would be interesting to explore which other variables and theories can exhibit their influence on purchase decision of discounted search goods under time pressure. Third, our measures of reliance of review valence had excellent psychometric properties; they did not address the issues surrounding the construct and items scales. Researchers can pursue the development of measures to better represent the constructs.

# 9. APPENDIX

| Items | Std. loadings | Composite reliability | Average variance extracted | Cronbach's alpha |
|---|---|---|---|---|
| **Reliance on review valence** | | 0.88 | 0.68 | 0.86 |
| If I do not consider aggregate rating before buying a product, I worry about my decision | 0.86 | | | |
| Aggregate ratings are more valuable to me than the opinion of my friends | 0.92 | | | |
| I trust aggregate ratings more than the opinion of those around me | 0.72 | | | |
| **Attitude toward review valence** | | 0.86 | 0.66 | 0.88 |
| Online aggregate ratings are helpful for my decision-making | 0.88 | | | |
| Online aggregate ratings make me confident in purchasing a product | 0.67 | | | |
| I find online aggregate ratings to be informative | 0.58 | | | |
| Online aggregate ratings are a great way to discover good things about products and services | 0.78 | | | |
| Online aggregate ratings are a great way to discover bad things about products and services | 0.66 | | | |
| **Review credibility** | | 0.88 | 0.72 | 0.87 |
| Not dependable . . . Dependable | 0.61 | | | |
| Not trustworthy . . . Trustworthy | 0.84 | | | |
| Not credible . . . Credible | 0.72 | | | |
| Not believable . . . Believable | 0.91 | | | |
| Not reputable . . . Reputable | 0.82 | | | |
| **Review diagnosticity** | | 0.84 | 0.70 | 0.86 |
| I find individual review ratings to be informative | 0.68 | | | |
| I find in-depth and detailed reviews to be informative | 0.72 | | | |
| I find information in the reviews to be understandable and readable | 0.58 | | | |
| I find reviewers' profile to be authentic | 0.88 | | | |
| **Review effectiveness** | | 0.83 | 0.78 | 0.88 |
| I find the review helpful in making the purchase | 0.86 | | | |
| The information in the review motivates me to purchase the product | 0.72 | | | |
| I find the popular reviews to be very relevant with product information | 0.86 | | | |
| **Perceived uncertainty** | | 0.84 | 0.72 | 0.86 |
| I feel uncertain about the information in the review | 0.88 | | | |
| I feel uncertain about reviewers' experience with the product | 0.62 | | | |
| I feel uncertain about the authenticity of the aggregate ratings | 0.78 | | | |
| **Purchase decision** | | | | |
| I would like to purchase the product | Dependent variable | | | |

**Table 1. Measurement model analysis**

|   | 1 | 2 | 3 | 4 | 5 | 6 |
|---|---|---|---|---|---|---|
| 1. Reliance on review valence | **0.811** | | | | | |
| 2. Attitude toward review valence | 0.218 | **0.834** | | | | |
| 3. Review credibility | 0.446 | 0.328 | **0.868** | | | |
| 4. Review diagnosticity | 0.403 | 0.160 | 0.172 | **0.824** | | |
| 5. Review effectiveness | 0.268 | 0.327 | 0.228 | 0.162 | **0.812** | |
| 6. Perceived uncertainty | 0.116 | 0.186 | 0.366 | 0.432 | 0.436 | **0.812** |

Notes: Diagonals represent the square root of the AVE

**Table 2. Results of tests for discriminant validity of study constructs**

| | Descriptive statistics | | One sample t-test | |
|---|---|---|---|---|
| **Constructs** | **Mean** | **SD** | **$t$ (df)** | **$p$** |
| Reliance on review valence | 4.32 | 1.20 | 4.48 (282) | 0.008 |
| Attitude towards review valence | 5.27 | 1.51 | 3 (282) | 0.000 |
| Review credibility | 5.67 | 1.78 | 1.65 (298) | 0.000 |
| Review diagnosticity | 4.82 | 1.13 | 4.66 (271) | 0.001 |
| Review effectiveness | 5.69 | 1.61 | 1.94 (271) | 0.000 |
| Perceived uncertainty | 4.89 | 1.16 | 13 (288) | 0.000 |
| Purchase decision | 5.74 | 1.32 | 9.41 (282) | 0.000 |

**Table 3. Descriptive statistics**

| | Hypotheses | | Estimate | S.E. | C.R | P value | Results |
|---|---|---|---|---|---|---|---|
| **H1** | Reliance on Review valence | → Purchase decision | 0.42 | 0.112 | 2.842 | 0.000* | Supported |
| **H2** | Attitude towards Review valence | → Reliance on Review valence | 0.68 | 0.162 | 1.432 | 0.000* | Supported |
| **H3** | Review credibility | → Attitude towards Review valence | 0.76 | 0.132 | 2.682 | 0.000* | Supported |
| **H4** | Review diagnosticity | → Attitude towards Review valence | 0.48 | 0.174 | 1.786 | 0.004* | Supported |
| **H5** | Review effectiveness | → Attitude towards Review valence | 0.52 | 0.156 | 2.016 | 0.000* | Supported |

*p-value<0.01

**Table 4. Result of hypothesized structural model**

| **Hypothesis** | **Constrained model** | **Unconstrain-ed model** | **Chi-square difference** | **Result on moderati-on** | **Result on hypothesis** |
|---|---|---|---|---|---|
| H6   Perceived uncertainty moderation effect | 388.486 (df = 282) | 369.320 (df = 278) | 20.368 | Significant | Supported |

**Table 5. Result of the effects of moderating variables**